\documentclass[prl,twocolumn,showpacs]{revtex4}

\usepackage{graphicx}
\usepackage{epsfig}
\usepackage{amsfonts}

\begin{document}
\title{Enhanced shot noise in resonant tunnelling via interacting localised states}
\author{S.S. Safonov$^1$, A.K. Savchenko$^1$, D.A. Bagrets$^2$, O.N. Jouravlev$^2$, Y.V. Nazarov$^2$, E.H. Linfield$^3$, and D.A. Ritchie$^3$}
\affiliation{$^1$ School of Physics, University of Exeter, Stocker Road, Exeter, EX4 4QL, United Kingdom\\
$^2$ Department of Applied Physics, Delft University of Technology, Lorentzweg 1, 2628 CJ Delft, The Netherlands\\
$^3$ Cavendish Laboratory, University of Cambridge, Madingley Road, Cambridge, CB3 0HE, United Kingdom}

\begin{abstract}
In a variety of mesoscopic systems shot noise is seen to be
suppressed in comparison with its Poisson value. In this work we
observe a considerable enhancement of shot noise in the case of
resonant tunnelling via localised states. We present a model of
correlated transport through two localised states which provides
both a qualitative and quantitative description of this effect.
\end{abstract}


\maketitle

Understanding the role of electron coherence and Coulomb
interaction in electron transport is one of the main directions of
contemporary research in mesoscopic physics. Recently, shot noise
measurements have proved to be a useful tool for these studies,
since they provide information which is not available from
standard conductance measurements \cite{Blanter!Buttiker!Review}.
Shot noise, i.e. fluctuations of the current in time due to the
discrete nature of electrons, is a measure of temporal
correlations between individual electron transfers through a
mesoscopic system. Uncorrelated transfers result in the Poisson
shot noise with the noise power $S_I=2eI$ ($e$ is the electron
charge, and $I$ is the average current). The effects on noise of
the Pauli exclusion principle \cite{Les} and the Coulomb repulsion
\cite{Korotkov} turn out to be similar in most mesoscopic systems.
Both were predicted to impose a time delay between two consecutive
electron transfers, which results in negative correlations between
them and, therefore, suppression of shot noise. This idea has been
intensively explored in studies of the shot noise properties in
ballistic and diffusive systems \cite{QPC,One!Third}.

\begin{figure}[b]
\psfig{figure=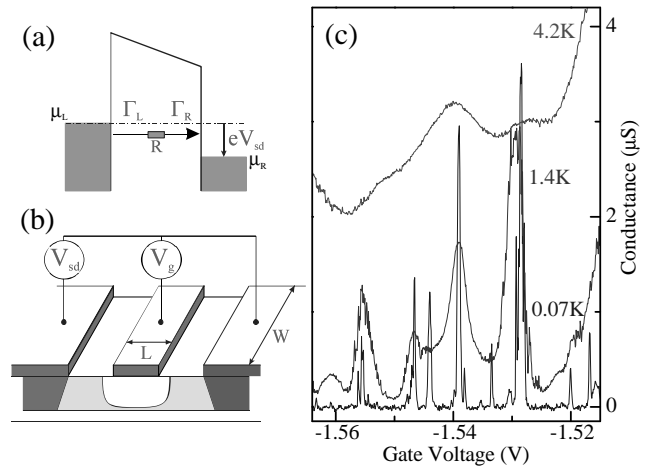,width=0.47\textwidth} \caption{(a) Resonant tunnelling through a localised state in
a barrier. (b) Cross-section of the transistor structure with two ohmic contacts and the gate between them.
(c) Typical RT peaks in the ohmic conductance at different temperatures.} \label{Peaks}
\end{figure}

Electron transport via localised states in a potential barrier
between two contacts has been a subject of intensive
investigations. If the size of a mesoscopic barrier is small,
resonant tunnelling (RT) through a single localised state
(impurity) becomes responsible for conduction across the barrier
\cite{Fowler}. When the resonant level ($R$) coincides with either
of the Fermi levels in the contacts, $\mu _{L,R}$, a peak in the
conductance appears. The amplitude of the peak is determined by
the ratio of the leak rates $\Gamma _{L,R}\propto \exp \left(
-2r_{L,R}/a\right)$ from the resonant impurity to the contacts,
where $r_{L,R}$ are the distances between the impurity and the
left or right contacts, and $a$ is the localisation radius of the
state, Fig.\ref{Peaks}a. The current is given by the relation
$I_0= e \Gamma _L\Gamma _R/\left( \Gamma _L+\Gamma _R \right)$. It
has been predicted in \cite{Ting,Nazarov} that for RT via a
localised state shot noise is suppressed by the Fano factor $F
\equiv S_I/2eI_0 = \left( \Gamma _L^2+\Gamma _R^2 \right)/\left(
\Gamma _L+\Gamma _R\right) ^2$. The Fano factor then ranges from
0.5 (for equal rates) to 1 (for significantly different rates)
dependent on the position of the resonant impurity inside the
barrier. Suppression of shot noise in accordance with this
relation has been first observed in a resonant tunnelling
structure \cite{Old}. Similar suppression of shot noise in the
Coulomb blockade regime has been seen in a quantum dot
\cite{Schonenberger}. In electron hopping (sequential tunnelling)
through $N$ equivalent barriers the Fano factor is also expected
to be suppressed, $F=1/N$, if one assumes that the Poisson noise
is generated across a single barrier \cite{Kuznetsov!Hopp}.

In this work we present a study of time-dependent fluctuations of
the RT current through a short (0.2 $\mu$m) tunnel barrier.
Surprisingly, we observe a significant \emph{enhancement} of shot
noise with respect to the Poisson value. We explain this effect by
correlated resonant tunnelling involving two interacting localised
states.

The experiment has been carried out on a \emph{n}-GaAs MESFET
consisting of a GaAs layer of $0.15$ $\mu$m (donor concentration
$10^{17}$ cm$^{-3}$) grown on an undoped GaAs substrate. On the
top of the structure an Au gate is deposited with dimensions
$L=0.2$ $\mu $m in the direction of the current and $W=20$ $\mu $m
across it, Fig.\ref{Peaks}b. By applying a negative gate voltage,
$V_g$, a lateral potential barrier is formed between the ohmic
contacts (source and drain).  Its height is varied by changing
$V_g$. When a source-drain voltage $V_{sd}$ is applied,
fluctuations of the current between the ohmic contacts are
measured by two low-noise amplifiers. The cross-correlation
spectrum in the frequency range $50-100$ kHz is detected by a
spectrum analyzer \cite{Roshko}. This technique removes noise
generated by the amplifiers and leads.

Fig.\ref{Peaks}c shows an example of conductance peaks as a
function of $V_g$ in the studied sample at $V_{sd}=0$ and
different temperatures down to $T=0.07$ K. One can see that with
lowering temperature the background conduction (due to electron
hopping) decreases and the amplitude of the conductance peaks
increases. This increase is a typical feature of resonant
tunnelling through an impurity \cite{Fowler}.

\begin{figure}[t]
\psfig{figure=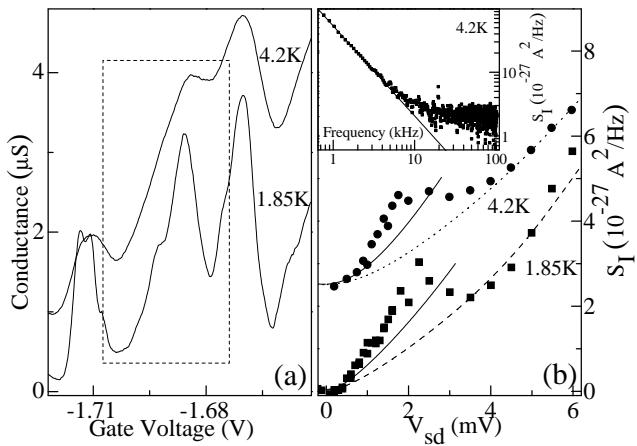,width=0.47\textwidth} \caption{(a) Conductance peaks in the region of $V_g$ where
the current noise has been measured. (b) Shot noise power as a function of $V_{sd}$: at $V_g=-1.6945$ V for
$T=1.85$ K and $V_g=-1.696$ for $T=4.2$K. Lines show the dependences $S_I\left( V_{sd}\right) $ expected for
resonant tunnelling through a single impurity from Eq.(\ref{Fit}), with $F=1$ (solid), $F=0.63$ (dashed), and
$F=0.52$ (dotted). Inset: Excess noise spectrum at $V_g=-1.696$ V and $V_{sd}=1.5$ mV.} \label{Noise1}
\end{figure}

The box in Fig.\ref{Noise1}a indicates the  range of $V_g$ where
shot noise has been studied at 1.85 K$<T<4.2$ K. In
Fig.\ref{Noise1}b (inset) an example of the excess noise spectrum
is shown at a gate voltage near the RT peak in Fig.\ref{Noise1}a.
(In this spectrum thermal noise has been subtracted and the effect
of the stray capacitance has been taken into account according to
\cite{Roshko}.) Shot noise is determined from the flat region of
the spectrum above 40 kHz. In this region one can neglect the
contribution of $1/f^\gamma $ noise ($\gamma =1.6$) which is shown
in Fig.\ref{Noise1}b (inset) by a solid line.

Fig.\ref{Noise1}b shows the dependence of the shot noise power on
$V_{sd}$ at two temperatures. At small biases ($V_{sd}<3$ mV) a
pronounced peak in noise is observed with an unexpectedly large
Fano factor $F>1$. This is seen by plotting the dependences
$S_I\left( V_{sd}\right) $ with different $F$ using the
phenomenological expression for excess noise in the case of RT
through a single impurity (cf. Eq.(62) in
\cite{Blanter!Buttiker!Review} and Eq.(11) in \cite{Nazarov}):
\begin{equation} S_I=F2eI_{sd}\coth \left(
\frac{eV_{sd}}{2k_BT}\right) -F4k_BTG_S. \label{Fit}
\end{equation}
(The expression describes the evolution of excess noise into shot
noise $S_I=F2eI_{sd}$  at $eV_{sd}>k_BT$; $G_S$ is the ohmic
conductance of the sample) At large biases ($V_{sd}>3$ mV),
however, shot noise decreases to a conventional sub-Poisson value,
$F\sim 0.6$.

We have established that the increase of shot noise exists only in
a specific range of $V_g$. It is worth noting that there is no
negative differential conductance in the region of $V_{sd}$-$V_g$
where the peak in the noise appears, and, therefore, we cannot
link this enhancement to some sort of instability
\cite{Blanter!Buttiker!NDC}. Instead, we will show that in this
region of $V_{sd}$-$V_g$ the resonant current is carried by
\emph{two interacting impurities} and this leads to the increase
of shot noise.

\begin{figure}[t]
\psfig{figure=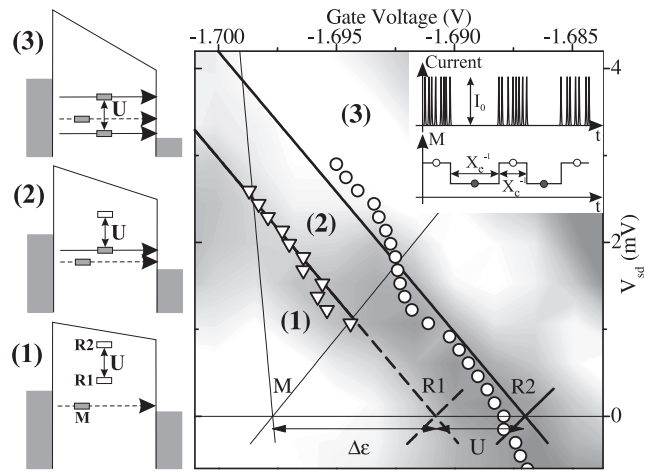,width=0.47\textwidth} \caption{Left panel: Energy diagrams of the two impurities
for different positive $V_{sd}$: $V_{sd}^{(1)}<V_{sd}^{(2)}<V_{sd}^{(3)}$. Inset: Schematic representation of
the modulation of the current through impurity $R$ by changing the occupancy of modulator $M$. Main part:
Grey-scale plot of the differential conductance as a function of $V_{g}$ and $V_{sd}$ at $T=1.85$ K (darker
regions correspond to higher differential conductance, background hopping contribution is subtracted). Lines
show the positions of the conductance peaks of impurity $R$ and modulator $M$ obtained from the fitting of
the noise data in Fig.\ref{Noise2}.} \label{Grey}
\end{figure}

We will first show that interaction between two states can
considerably increase shot noise. Let us start with a simple
illustrative model and consider two spatially close impurity
levels, $R$ and $M$, separated in the energy scale by $\triangle
\epsilon$. If impurity $M$ is charged, the energy level of $R$ is
shifted upwards by the Coulomb energy $U\sim e^2/\kappa r$, where
$r$ is the separation between the impurities and $\kappa $ is the
dielectric constant, Fig.\ref{Grey} (diagram 1). Thus, dependent
on the occupancy of $M$, impurity $R$ can be in two states: $R1$
or $R2$. Further we assume that $V_{sd}$ is small enough so that
state $R2$ is above the Fermi level in the left contact,
Fig.\ref{Grey} (diagram 2). Then electrons are transferred via $R$
with the rates $\Gamma_{L,R}$ if $M$ is empty, and \emph{cannot}
be transferred if $M$ is charged. It is assumed that impurity $M$
changes its states independently of the state of impurity $R$:
from empty to \textit{charged} state with the rate $X_c$ and from
charged to \textit{empty} state with the rate $X_e$. If $M$
changes its occupancy at a slow rate, i.e. $X_{e,c}\ll \Gamma
_{L,R}$, its contribution to the current is negligible and we will
call $M$ a \emph{modulator} since it modulates the current through
impurity $R$. This current jumps randomly between two values:
zero, when $M$ is occupied, and $I_0$ when $M$ is empty,
Fig.\ref{Grey} (inset). If the bias is increased, the upper state
$R2$ is shifted down into the conducting energy strip and the
modulation of the current via impurity $R$ vanishes,
Fig.\ref{Grey} (diagram 3).

In the modulation regime, the average current through impurity $R$
and the corresponding zero-frequency Fano factor can be written,
respectively, as
\begin{equation}
I=\frac{e\Gamma _L\Gamma _R}{\Gamma _L+\Gamma _R}\frac{X_e}{X_e+X_c} \label{current}
\end{equation}
and
\begin{equation}
F=\frac{\Gamma _L^2+\Gamma _R^2}{\left(\Gamma _L+\Gamma _R\right)^2}+ 2\frac{\Gamma _L\Gamma _R}{\Gamma
_L+\Gamma _R}\frac{X_c}{\left(X_e+X_c\right)^2}. \label{Fano}
\end{equation}

The first term in Eq.(\ref{Fano}) describes the suppression of the
Fano factor below unity \cite{Nazarov}, whereas the second term
gives a positive contribution. To illustrate the origin of the
second term, one can think of the modulated current as random
telegraph noise (RTN), i.e. spontaneous jumps between zero and
$I_0$. The second term can then be obtained from the spectrum of
RTN \cite{Kirton!Uren} with characteristic times of the upper and
lower states - $1/X_e$ and $1/X_c$, respectively. If $X_{e,c}\ll
\Gamma _{L,R}$, a substantial enhancement of shot noise, $F \gg
1$, is expected from Eq.(\ref{Fano}). Another way to illustrate
the origin of this effect is to assume that $M$ is close to the
left contact. As a result, impurity $M$ spends more time in its
charged state, i.e. $X_e\ll X_c$ and the current through $R$ is
transferred in bunches, with the average duration of a bunch
$\tau_e=1/X_c$. The noise due to the `chopping' of the current can
then be estimated as $S_I=2QI$, where $Q$ is the average charge
transferred in one bunch. This charge is equal to $I_0\tau_e$, and
this again gives the second term in Eq.(\ref{Fano}).

This model of a slow modulator which changes its state
independently of impurity $R$ may look too simplistic. However,
its generalisation (for any relation between $X$ and $\Gamma $) is
straightforward and provides a consistent quantitative description
of the observed effect. Our theoretical model is based on the
master equation formalism \cite{Nazarov,Glazman!Matveev}. It is
applicable when $\Gamma_{L,R} <k_BT$ - the condition satisfied in
our experiment. Then the system of two interacting impurities $R$
and $M$ can be in four possible states. The transition rates
between these states are determined by tunnelling between the
contacts and impurities and depend on temperature and the level
positions with respect to the Fermi levels $\mu_{L,R}$. The
resulting transport problem is reduced to numerical
diagonalisation of a $4 \times 4$ matrix. As a result, the current
and the Fano factor are obtained as a function of the energy
positions of the two impurities, which are linearly dependent on
$V_{sd}$ and $V_g$. It is important that in our calculations the
effect of temperature, which suppresses the enhanced Fano factor
in Eq.(\ref{Fano}), is taken into account. (In a similar master
equation approach an increase of shot noise for two interacting
quantum dots was also obtained in \cite{Eto}, however at $T=0$.)

By measuring the differential conductance as a function of $V_g$
and $V_{sd}$ we have been able to show directly that the increase
of shot noise occurs in the region of $V_g$-$V_{sd}$ where two
interacting impurities carry the current in a correlated way.
Fig.\ref{Grey} presents the grey scale of the differential
conductance plotted versus $V_g$ and $V_{sd}$. When a source-drain
bias is applied, a single resonant impurity would give rise to two
peaks in $dI/dV\left( V_g\right) $, which occur when the resonant
level aligns with the Fermi levels $\mu_{L,R}$. On the grey scale
these peaks lie on two lines crossing at $V_{sd}=0$. Consider, for
example, point $M$ in Fig.\ref{Grey}. The central area between the
lines corresponds to the impurity level between $\mu_{L}$ and
$\mu_{R}$, that is when the impurity is in its conducting state.
Outside this region the impurity does not conduct, as it is either
empty (on the left of the central region) or filled (on the right
of it).

Experimentally, at small $V_{sd}$ we see such a cross-like feature
near point $R2$, with the left line being more pronounced. The
exact positions of the maxima of the conductance peaks of this
line are indicated by circles. It is seen that with increasing
$V_{sd}$, a new parallel line $R1$ appears at $V_g \approx -1.694$
V and $V_{sd} \approx 1$ mV, shifted to the left by $\triangle V_g
\approx 4$ mV. The maxima of the conductance peaks of this line
are shown by triangles.

In Fig.\ref{Grey} the modulator cross is plotted according to the analysis below - experimentally we cannot
observe these lines because the modulator conductance peaks are too small, due to low leak rates $X_e$ and
$X_c$. The $R1$-line occurs in the inner region of the modulator cross, i.e. where the modulator occupancy
changes in time. Therefore, lines $R1$ and $R2$ reflect the Coulomb shift of level $R$: the former
corresponds to the empty modulator and the latter - to the occupied one. The modulation of the current should
then occur in region (2), Fig.\ref{Grey}: the central part of cross $M$ between lines $R1$ and $R2$,
corresponding to diagram (2). As discussed before, in region (3) there is no modulation as both states $R1$
and $R2$ can conduct, and in region (1) there is no current as the low state $R1$ is still above $\mu _L$.

\begin{figure}[t]
\psfig{figure=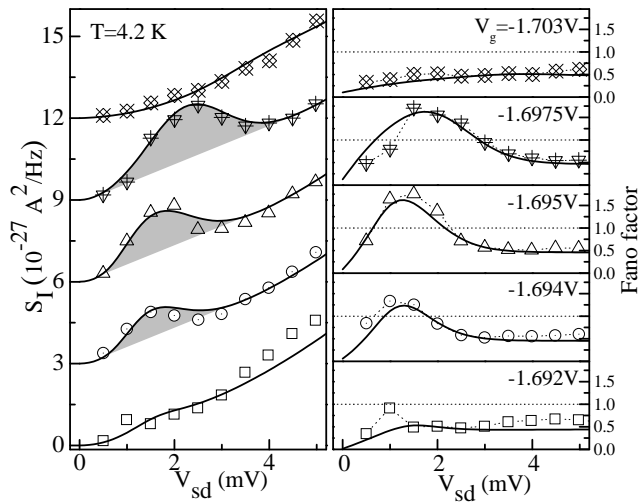,width=0.47\textwidth} \caption{Shot noise and the corresponding Fano factor as a
functions of source-drain bias at different gate voltages. (The shot noise data for different $V_g$ are
offset for clarity.) Solid lines show the results of the numerical calculations.} \label{Noise2}
\end{figure}

In Fig.\ref{Noise2} current noise and the Fano factor are presented as functions of $V_{sd}$ for different
$V_g$. It shows that indeed the increase of noise occurs only in region (2) in Fig.\ref{Grey}. Namely, the
increase of noise appears only between $V_g=-1.699$ V and $V_g=-1.693$ V, that is, in the central region of
cross $M$. In addition, when $V_{sd}$ is swept at fixed $V_g$, one can see that the hump in the Fano factor
appears only between lines $R1$ and $R2$.

In order to quantitatively compare the model with the experiment
we have to take into account that in our experiment resonant
tunnelling via state $R$ exists in parallel with the background
hopping. Then the total Fano factor has to be expressed as
$F=\left( F_{RT}I_{RT}+F_{B}I_{B}\right) /\left(
I_{RT}+I_{B}\right)$, where $F_{RT}$, $F_B$ and $I_{RT}$, $I_B$
are the Fano factors and currents for RT and hopping,
respectively. In order to get information about the background
hopping we have measured noise at $V_g>-1.681$ V, i.e. away from
the RT peak under study in Fig.\ref{Grey}. It has been estimated
as $F_B\sim 0.4$. This value of the Fano factor is expected for
shot noise in hopping through $N\sim 2-3$ potential barriers (1-2
impurities in series) \cite{Roshko,Korotkov!Likharev1D}). The bias
dependence of the background current at this $V_g$ is also
consistent with hopping current via two impurities: $dI/dV\propto
V_{sd}^{4/3}$ \cite{Glazman!Matveev}.

Assuming that the background current is approximately the same for
all studied gate voltages, we have added up the contributions to
the Fano factor from RT via two interacting impurities, $R$ and
$M$, and the background hopping. The numerical results have been
fitted to the experimental $dI/dV\left( V_{sd},V_g\right) $ and
$S_I \left( V_{sd}, V_g\right) $, Fig.\ref{Noise2}. The fitting
parameters are the leak rates of $R$ and $M$ ($\hbar \Gamma
_{L}\simeq 394$ $\mu $eV, $\hbar \Gamma _{R}\simeq 9.8$ $\mu $eV,
and $\hbar X_e\simeq 0.08$ $\mu $eV, $\hbar X_c\simeq 0.16$ $\mu
$eV ), the energy difference between $R$ and $M$ ($\triangle
\varepsilon =1$ meV), and the Fano factor for the background
hopping ($F_B=0.45$). The coefficients in the linear relation
between the energy levels $M$, $R$ and $V_{sd}$, $V_g$ have also
been found to match both the experimental data in Fig.\ref{Noise2}
and the position of lines $R1$ and $R2$ in Fig.\ref{Grey}. One can
see that the model gives good agreement with the experiment. The
Coulomb shift ($U\sim 0.55$ meV) found from Fig.\ref{Grey} agrees
with the estimation for the Coulomb interaction between two
impurities not screened by the metallic gate: $U\sim e^2/\kappa d
\sim 1$ meV, where $d\sim 1000$ \AA \, is the distance between the
gate and the conducting channel.

It is interesting to note that the hopping background effectively
hampers the manifestation of the enhanced Fano factor $F_{RT}$,
i.e. without the background the Fano factor enhancement would be
much stronger. The largest experimental value of $F$ in
Fig.\ref{Noise2} (at $V_g=-1.6975$ V) is approximately $1.5$,
while a numerical value for RT at this $V_g$ is $F_{RT} \approx
8$.

In conclusion, we have observed enhanced shot noise in resonant
tunnelling via localised states in a short-barrier structure. We
have demonstrated that this effect originates from Coulomb
interaction between two localised states which imposes
correlations between electron transfers. A simple model is shown
to provide a quantitative description of the observed enhancement.

We are grateful to E.V. Sukhorukov for stimulating discussions,
FOM, EPSRC and ORS fund for financial support.

\end{document}